\documentclass[aps,prl,twocolumn,showpacs,superscriptaddress,nofootinbib,preprintnumbers]{revtex4-2}

\usepackage{style}
\usepackage{lineno}

\newcommand{\causalacronym}{ECT}
\newcommand{\rc}{\ensuremath{r_{\rm ect}}}
\begin{document}
\vspace{-0.5cm}
\hfill{\small \tt ~~~~~~~~~~~~~ FERMILAB-PUB-25-0523-T}
\hfill
\vspace{0.5cm}


\title{\textbf{First constraints on causal sources of primordial gravitational waves \\ from BICEP/Keck, SPTpol, SPT-3G, Planck and WMAP $B$-mode data}}

\author{Jessica A. Zebrowski\,\orcidlink{0000-0003-2375-0229}} 
\email{j.z@uchicago.edu}
\affiliation{Fermi National Accelerator Laboratory, Batavia, Illinois 60510}
\affiliation{Kavli Institute for Cosmological Physics, University of Chicago, Chicago, IL 60637}

\author{Aurora Ireland\,\orcidlink{0000-0001-5393-0971}} 
\email{anireland@stanford.edu}
\affiliation{Leinweber Institute for Theoretical Physics, Stanford University, Stanford, CA 94305}

\author{Christian L. Reichardt\,\orcidlink{0000-0003-2226-9169}}
\email{christian.reichardt@unimelb.edu.au}
\affiliation{School of Physics, University of Melbourne, Parkville, VIC 3010, Australia}

\author{Kylar Greene\,\orcidlink{0000-0002-2711-7191}}
\email{kylar.cosmo@pm.me}
\affiliation{Department of Physics and Astronomy, Seoul National University, 1 Gwanak-ro, Gwanak-gu, Seoul 08826, Korea}

\author{Gordan Krnjaic} 
\email{krnjaicg@uchicago.edu}
\affiliation{Fermi National Accelerator Laboratory, Batavia, Illinois 60510}
\affiliation{Kavli Institute for Cosmological Physics, University of Chicago, Chicago, IL 60637}
\affiliation{Department of Astronomy and Astrophysics, University of Chicago, Chicago, IL 60637}

\author{Yuhsin Tsai\,\orcidlink{0000-0001-7847-225X}}
\email{ytsai3@nd.edu}
\affiliation{Department of Physics and Astronomy, University of Notre Dame, South Bend, IN 46556}

\author{Fran\c{c}ois R Bouchet\,\orcidlink{0000-0002-8051-2924}}
\affiliation{Sorbonne Universit\'{e}, CNRS, UMR 7095, Institut d’Astrophysique de Paris, 98 bis bd Arago, 75014 Paris, France}

\date{\today}

\newcommand{\rlimit}{\ensuremath{0.033}}
\newcommand{\sigmar}{\ensuremath{0.011}}
\newcommand{\aclimit}{\ensuremath{0.0077}}
\newcommand{\sigmaac}{\ensuremath{0.0014}}
\newcommand{\alenslimit}{\ensuremath{1.01^{+0.02}_{-0.02}}}
\newcommand{\sigmaalens}{\ensuremath{0.022}}

\newcommand{\rlimitThreeGSim}{\ensuremath{0.01}}
\newcommand{\sigmarThreeGSim}{\ensuremath{0.0033}}
\newcommand{\aclimitThreeGSim}{\ensuremath{0.00059}}
\newcommand{\sigmaacThreeGSim}{\ensuremath{0.00022}}
\newcommand{\alenslimitThreeGSim}{\ensuremath{-}}
\newcommand{\sigmaalensThreeGSim}{\ensuremath{-}}

\newcommand{\rlimitThreeGDelensSim}{\ensuremath{0.0047}}
\newcommand{\sigmarThreeGDelensSim}{\ensuremath{0.0015}}
\newcommand{\aclimitThreeGDelensSim}{\ensuremath{0.0003}}
\newcommand{\sigmaacThreeGDelensSim}{\ensuremath{0.00012}}
\newcommand{\alenslimitThreeGDelensSim}{\ensuremath{-}}
\newcommand{\sigmaalensThreeGDelensSim}{\ensuremath{-}}

\newcommand{\rlimitThreeGPlusSim}{\ensuremath{0.01}}
\newcommand{\sigmarThreeGPlusSim}{\ensuremath{0.0032}}
\newcommand{\aclimitThreeGPlusSim}{\ensuremath{0.00029}}
\newcommand{\sigmaacThreeGPlusSim}{\ensuremath{0.00011}}
\newcommand{\alenslimitThreeGPlusSim}{\ensuremath{-}}
\newcommand{\sigmaalensThreeGPlusSim}{\ensuremath{-}}

\newcommand{\rlimitThreeGPlusDelensSim}{\ensuremath{0.0011}}
\newcommand{\sigmarThreeGPlusDelensSim}{\ensuremath{3.4\times 10^{-4}}}
\newcommand{\aclimitThreeGPlusDelensSim}{\ensuremath{1.1\times 10^{-4}}}
\newcommand{\sigmaacThreeGPlusDelensSim}{\ensuremath{2.5\times 10^{-5}}}
\newcommand{\alenslimitThreeGPlusDelensSim}{\ensuremath{-}}
\newcommand{\sigmaalensThreeGPlusDelensSim}{\ensuremath{-}}


\begin{abstract}
Non-inflationary sources of gravitational waves in the early Universe generically predict causality-limited tensor power spectra at low frequencies. 
We report the first-ever constraints on such sources based on cosmic microwave background (CMB) $B$-mode polarization measurements.
Using data from BICEP/Keck, SPTpol, SPT-3G, \textit{Planck}, and \textit{WMAP}, we constrain the amplitude of an early causal tensor (ECT) power spectrum parameterized by \rc{}, the ratio of causal tensor power to total scalar power at $k~=~0.01$\,Mpc$^{-1}$, and obtain a 95\% CL upper limit of $\rc{} <$ \aclimit{}. 
Since \rc{} can easily be related to the  parameters of a given theory, our bound robustly constrains a broad class of well-motivated gravitational wave sources in the early universe, including first-order cosmological phase transitions, enhanced small-scale density perturbations, and various topological defects. 
Finally, we translate our limit into a  bound on the present-day energy density in gravitational waves at ultra-low frequencies otherwise inaccessible to traditional gravitational wave detection strategies, including pulsar timing arrays, interferometers, and resonant cavities. 

\end{abstract}

\maketitle

\section{Introduction}

Tensor perturbations in the early universe imprint a distinctive $B$-mode pattern in the polarization of the cosmic microwave background (CMB), offering a powerful observational probe of fundamental physics at energy scales beyond the reach of 
terrestrial gravitational wave (GW) experiments~\cite{kamionkowski1997probe}.
While this connection has long been appreciated in the context of inflation, it has only recently been recognized that non-inflationary sources of primordial tensor perturbations can also yield observable $B$ modes, potentially complicating interpretations of future measurements~\cite{Greene:2024,Ireland:2025,Sousa:2015}. Examples of such sources include first-order cosmological phase transitions~\cite{Kamionkowski:1993,Caprini:2015}, enhanced scalar density perturbations~\cite{Domenech:2021}, dark-photon perturbations from axion rolling~\cite{Geller:2021obo}, primordial gravitational vector modes~\cite{ali}, and various topological defects~\cite{Kibble:1976,Pogosian:2007gi,Vachaspati:1984,Hindmarsh:1994}. 

Unlike inflation, which stretches sub-horizon tensor perturbations to super-horizon scales and predicts a nearly scale-invariant power spectrum, post-inflationary sources 
are fundamentally limited by the impossibility of super-horizon correlations, and 
therefore universally yield white-noise spectra on sufficiently large scales. For all such sources, the dimensionless tensor power spectrum exhibits a characteristic $\mathcal{P}_h(k) \propto k^3$ scaling at small comoving wavenumber $k$. This behavior is a robust, model-independent consequence of causality and finite correlation lengths, and it holds universally for length scales larger than the physical extent of the source, provided that the latter is bounded in frequency and time~\cite{Cai:2019}. 
 

The characteristic $k^3$ scaling will turn over at some $k$ related to the correlation length of the source. 
However if the tensor source is before $z \sim 7 \times 10^4$, this turn over will be hidden in the CMB by Silk damping -- the horizon size at $z \sim 7 \times 10^4$ matches the Silk damping scale.
Any non-inflationary source that generates GWs exclusively before this time necessarily predicts a white-noise power spectrum on CMB observable scales; this characteristic scaling yields a universal shape for the resulting $B$-mode angular power spectrum. 
We refer to such phenomena as early causal tensor (\causalacronym{}) sources.


Compared to inflationary predictions,  \causalacronym{} sources yield more power on small scales and less power on large scales.
This striking shape difference allows for  clear discrimination between the signals from inflation and \causalacronym{} sources, given sufficiently precise measurements across a range of angular scales. 
While the \causalacronym{} $B$-mode angular spectrum coincidentally resembles corresponding contributions from lensing, this degeneracy can be broken with external priors on lensing $B$ modes. Furthermore, since all \causalacronym{} sources predict the same $B$-mode spectral shape, the only parameter that distinguishes  sources in this universality class is the overall amplitude, which we parametrize as \rc{}, the ratio of \causalacronym{} tensor power to  measured scalar power at the reference comoving wavenumber $k_{\rm ref} = 0.01$\,Mpc$^{-1}$.



\begin{figure}[t!]
\hspace{-0.6cm}
\includegraphics[width=0.48\textwidth]{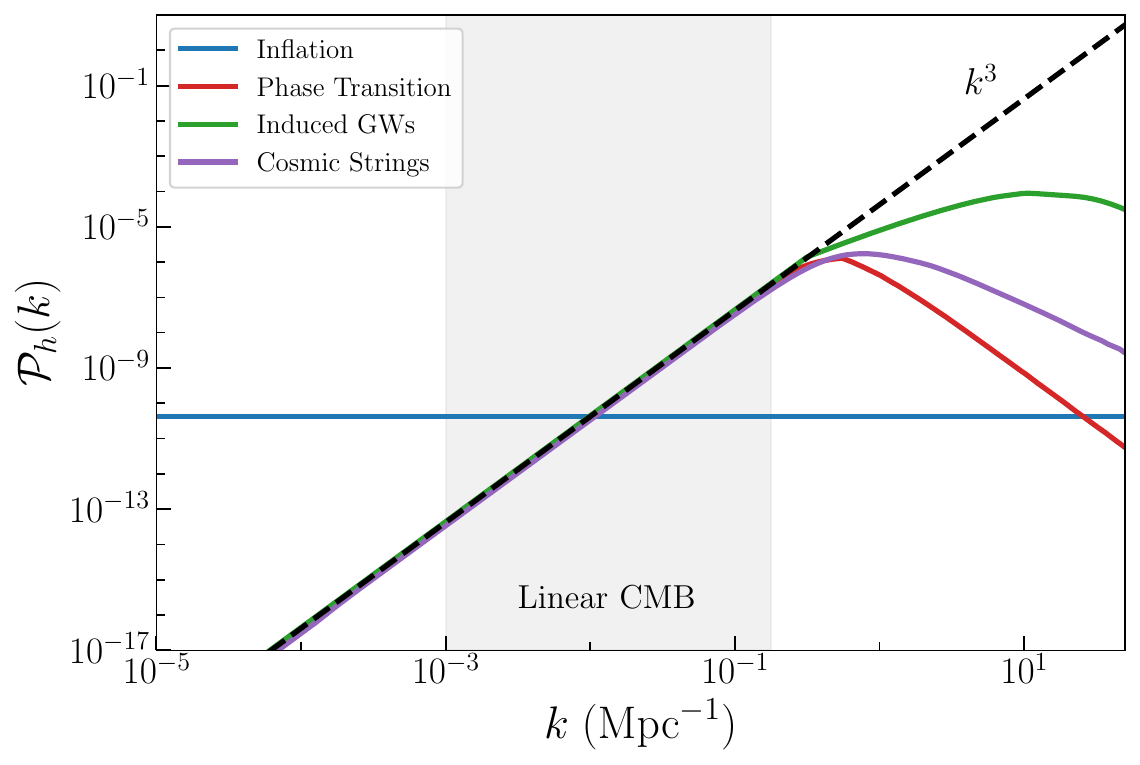}
\caption{Tensor power spectra $\mathcal{P}_h(k)$ for select ECT models presented by \cite{TheoryPaper}: a first-order cosmological phase transition (red), scalar-induced GWs (green), and cosmic strings (purple). The gray ``linear CMB'' region marks the domain of support of the window function $\mathcal{F}_\ell(k)$ in Eq. \eqref{eq:DlBB} for $20 < \ell < 2301$. 
As the models asymptote to the same $k^3$ scaling required by causality over the CMB regime, all three models predict identical $B$-mode power spectra. For comparison, we show  a scale-invariant inflationary spectrum (blue).
}
\label{fig:Ph}
\end{figure}

In this \emph{Letter}, we report the first constraints on the causal tensor power spectrum using contemporary CMB $B$-mode observations. 
A companion work~\cite{TheoryPaper} explores models which produce such causal tensors. 
Marginalizing over foreground and cosmological signals, we place the first upper limit on the amplitude of the causal tensor spectrum of $\rc{} <$\aclimit{} at 95\% CL. 
The limit constrains  a broad class of theoretical models that source GWs in the post-inflationary universe (see Fig.~\ref{fig:Ph}). 
We also translate this limit to a model-dependent bound on the GW energy density, $\Omega_{\rm GW}$, at ultra-low frequencies. 

\section{General Framework}\label{sec:theory}

We briefly review the ECT framework presented by \cite{TheoryPaper}, leaving details to that work.  
A Fourier mode of the transverse, traceless tensor perturbation $h_{ij}(\tau, \vec{k})$ evolves according to the wave equation
\begin{equation}\label{eq:fullwaveeq}
     h^{\prime \prime}_{ij} + 2 {\cal H} h^ \prime_{ij} + k^2 h_{ij} = 16 \pi G a^2 \, \Pi_{ij} \,,
\end{equation}
where $\tau$ is conformal time, $^\prime$ denotes differentiation with respect to $\tau$, $a$ is the scale factor, and ${\cal H} = a^\prime/a$ is the conformal Hubble rate. The source $\Pi_{ij}(\tau,\vec{k})$ is the transverse, traceless projection of the anisotropic stress tensor, whose functional form is model-dependent. We take $\Pi_{ij}$ to be bounded in frequency and non-vanishing only transiently in the early universe for some characteristic time interval $\tau \in [\tau_i, \tau_\star]$. 


We decompose  $h_{ij}$ and $\Pi_{ij}$ into the two polarizations $\lambda = (+,\times)$, which are uncorrelated and have the same spectra. 
The solutions can be written as $h^\lambda(\tau,\vec{k}) \equiv h^\lambda_\star(\vec{k}) \mathcal{T}(\tau,\vec{k})$ where $h^\lambda_\star(\vec{k}) \equiv h^\lambda(\tau_\star,\vec{k})$
is the initial amplitude, and $\mathcal{T}(\tau,\vec{k})$ is the tensor transfer function, which solves the sourceless version of Eq.~(\ref{eq:fullwaveeq}) and governs the deterministic evolution of these perturbations in an expanding universe. The initial amplitude depends on the source  and is obtained by solving Eq.~(\ref{eq:fullwaveeq}),
\begin{equation}
    h_\star^\lambda(\vec{k}) = 16 \pi G a^2 \int_{\tau_i}^{\tau_\star} d\tau' \, G_k(\tau,\tau') \Pi_\lambda(\tau',\vec{k}) ,
\end{equation}
where $G_k(\tau,\tau') = \sin[k(\tau - \tau')]/k$ is the Green's function for Eq.~\eqref{eq:fullwaveeq}, neglecting Hubble expansion over the short duration of source activity. Using this solution, we define the (\textit{dimensionless})  tensor power spectrum $\mathcal{P}_h$ through
\begin{equation}\label{eq:Ph}
    \left\langle h_\star^\lambda(\vec{k}) \, h_\star^{\lambda'}\!(\vec{k}')^* \right\rangle = \frac{\delta_{\lambda \lambda'}}{2} \frac{2\pi^2}{k^3} \mathcal{P}_h(k) (2\pi)^3 \delta^{(3)}(\vec{k} - \vec{k}') \,,
\end{equation}
where $\langle \dots \rangle$ denotes an ensemble average and $*$ is complex conjugation.

For any temporally-bounded source active on some characteristic sub-horizon scale $k_s$ during radiation domination, 
$\mathcal{P}_h(k) \propto k^3$ for all $k \ll k_s$~\cite{Cai:2019, TheoryPaper},
which reflects the causality requirement that sufficiently-separated spatial points must be uncorrelated. Thus, the power spectrum of any causal non-inflationary source must decay as white noise on super-horizon length scales. 
Given this, we parametrize the input power spectrum as 
\begin{equation}\label{eq:ansatz}
    \mathcal{P}_h(k) \equiv  \rc{}  A_s\left( \frac{k}{k_{\rm ref}} \right)^3 \,,
\end{equation}
where $k_{\rm ref}= 0.01$ Mpc$^{-1}$ is a reference wavenumber, $A_s = 2.1 \times 10^{-9}$ is the measured amplitude of scalar fluctuations~\cite{planck18-6}, and \rc{} is the amplitude of the white-noise tensor power spectrum at $k_{\rm ref}$ in units of $A_s$. 
This notation follows the long-standing convention to use the tensor-to-scalar ratio $r = A_T/A_s$ as a measure of inflationary tensor modes. 
For the chosen $k_{\rm ref} = 0.01$ Mpc$^{-1}$, the amplitude of the $B$-mode signals from inflationary and causal GWs approximately match at the recombination peak ($\ell \sim 100$) when  $r = \rc{}$.


%
To obtain the angular spectrum of $B$-mode polarization sourced by ${\cal P}_h$ in Eq.~\eqref{eq:ansatz}, we evaluate~\cite{Rubakov:2017}
\begin{equation}\label{eq:DlBB}
   {\cal D}_\ell^{\rm \causalacronym{}} =   18 \,\ell (\ell+1) \, T^2_0  \int_0^\infty  \frac{dk}{k} \mathcal{P}_h(k) \mathcal{F}_\ell(k)^2 \,,
\end{equation} 
where $T_0 = 2.725\, \text{K}$ is the CMB temperature today~\cite{Fixsen:2009ug} and $\mathcal{F}_\ell(k)$ is a source-independent window function defined in~\cite{Greene:2024}.

\begin{figure}
    \vspace{-0.25cm}
    \includegraphics[width=\linewidth]{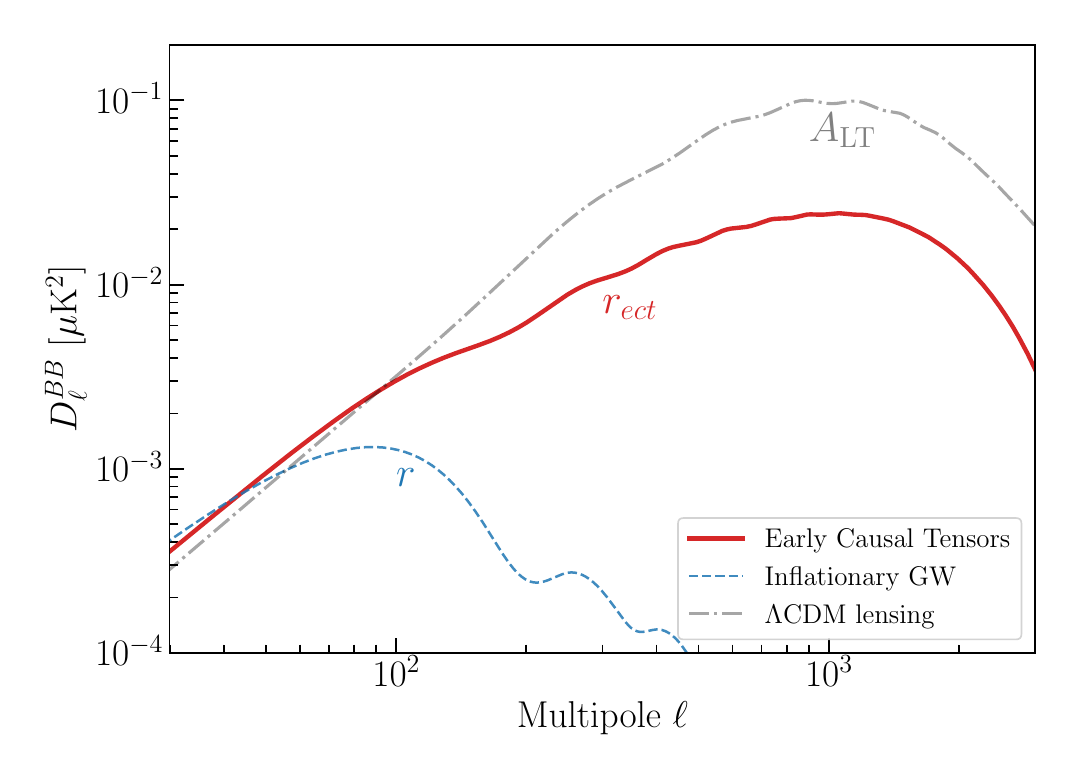}
    \caption{Theoretical $B$-mode power spectra for: inflationary GWs (blue dashed); ECT sources (red solid); and gravitational lensing (gray dot-dashed). The curves correspond to $r = 0.02$, $\rc{}= 0.02$, and $A_{\mathrm{LT}} = 1$. 
    }
    \label{fig:models}
\end{figure}

\section{Likelihoods}

We search for causal GWs using $B$-mode power spectrum measurements from BICEP/Keck~\cite{bicepkeck21c}, SPT-3G~\cite{zebrowski25}, and SPTpol~\cite{sayre20}. While these experiments have overlapping sky coverage, the SPT error bars are noise dominated. 
Thus, we treat them as independent when combining the likelihoods.
These three measurements are chosen as each has angular multipoles where they provide the tightest current statistical constraints.

\subsection{BICEP/Keck}

We use the BICEP/Keck \( B \)-mode likelihood presented in \citet{bicepkeck21c}, incorporating measurements through the 2018 observing season. The likelihood includes cross-spectra between all frequency bands from BICEP2, Keck Array (95, 150, and 220\,GHz), BICEP3 (95\,GHz), and external maps from \textit{Planck} (30--353\,GHz) and \textit{WMAP} (23--94\,GHz). The analysis jointly fits for CMB and foreground contributions, marginalizing over synchrotron and dust components. 
We use the public ``BK18'' likelihood, which combines the \( BB \) cross-spectra across all instruments to optimize foreground separation and maximize sensitivity to primordial \( B \) modes at large angular scales. 
These data currently yield the most precise measurement of the $B$-mode power spectrum at $\ell \lesssim  300$. 

\subsection{SPT-3G}

We use the SPT-3G $B$-mode likelihood presented by ~\citet{zebrowski25} for power spectrum measurements over the multipole range \( 32 < \ell < 502 \). These measurements are from 95, 150, and 220\,GHz observations of the SPT-3G 1500\,deg$^2$ patch observed from 2019-2020. 
Following the main cosmological results in~\cite{zebrowski25}, we focus on a subset of the sky overlapping with the BICEP/Keck survey area with the cleanest foregrounds. 
We use the same nuisance parameters as~\cite{zebrowski25}. 
These parameters include three absolute polarization calibration terms (one per frequency band), a Poisson term for radio galaxies with amplitude \( A^{\mathrm{95\,GHz,rg}}_{\ell=500} \), and three parameters for Galactic dust: the amplitude \(A^{\mathrm{150\,GHz}}_{\ell=80,\mathrm{SPT-3G}} \), angular power-law index \( \alpha_\mathrm{SPT-3G} \), and spectral index \( \beta_\mathrm{SPT-3G} \) of a modified blackbody spectrum.
The SPT-3G data provide the best measurement at $300 < \ell < 500$.

\subsection{SPTpol}

We use the SPTpol $B$-mode power spectrum measurement reported by \citet{sayre20}. 
The SPTpol bandpowers cover the angular multipole range $52\le \ell \le 2301$. 
While~\cite{sayre20} measure $B$-mode polarization in two bands, here we only use the higher signal-to-noise 150\,GHz power spectrum. 
We use the same four nuisance parameters as~\cite{sayre20}, including Galactic dust terms, Poisson radio galaxy emission, and  the absolute polarization calibration. 
We neglect the SPTpol beam uncertainty, given the modest signal-to-noise of the $B$-mode measurement. 
The SPTpol data provide the best $B$-mode measurement at $\ell > 500$.

\begin{figure}
    \begin{center}
        \hspace{-1cm}\includegraphics[width= \linewidth]{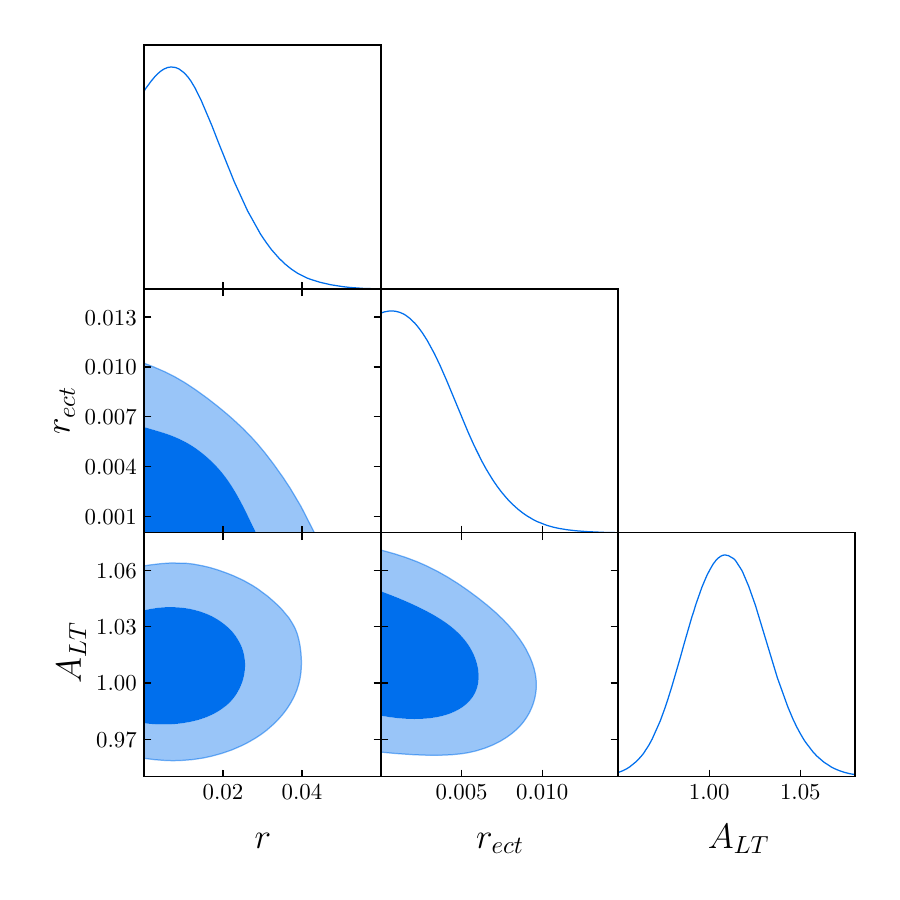}
        \caption{
        Constraints on the tensor-to-scalar ratio $r$, the ratio of causal tensor to scalar power \rc{}, and the amplitude of lensing $B$ modes $A_{\rm LT}$ from a combination of BICEP/Keck, SPTpol, SPT-3G, \textit{Planck}, and \textit{WMAP} data. The diagonal panels show the 1D posterior distributions of the parameters and the off-diagonal panels show the 2D 68\% and 95\% confidence intervals.}
        \label{fig:corner}
    \end{center}
\end{figure}

\begin{figure}
    \begin{center}
        \includegraphics[width=\linewidth]{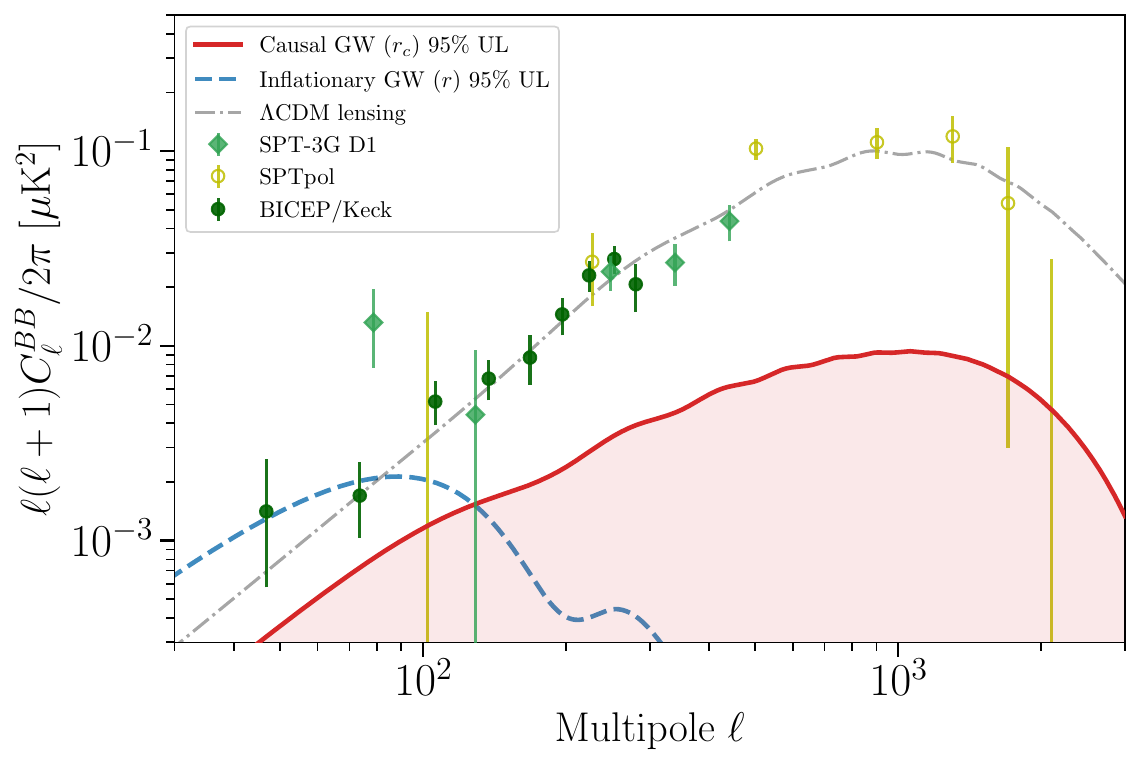}
        \caption{Measured \( B \)-mode power spectra from SPT-3G (diamonds), SPTpol (empty circles), and BICEP/Keck 2018 (filled circles), 
        plotted against the predicted lensing \( B \)-mode power with \( A_{\mathrm{LT}} = 1.03 \) (dash-dot grey line). The plotted SPT-3G \cite{balkenhol2024compressed} and BICEP/Keck points are CMB-only for visualization purposes; however parameter constraints use the full multi-frequency likelihood with foreground marginalization. We show model lines corresponding to the derived 95\% CL upper limits on the ECT (red) and inflationary (blue) GW power. The upper limit on ECT GWs is set by the absence of excess power at high multipoles, above that expected for lensing \( B \)-modes.
        }        
        \label{fig:ul}
    \end{center}
\end{figure}

\subsection{Model}\label{sec:Model}

We model the observed $B$-mode power spectrum as the sum of contributions from inflationary GW, \causalacronym{} sources, lensing-induced \( B \) modes, and frequency-dependent Galactic and extragalactic foregrounds,
\begin{equation}\label{eq:model} 
{\cal D}_\ell^{\rm BB} =  {\cal D}_\ell^{\mathrm{IGW}}(r) + {\cal D}_\ell^{\mathrm{\causalacronym{}}}(\rc{}) + A_{\mathrm{LT}} \, {\cal D}_\ell^{\mathrm{lens}} + {\cal D}_\ell^{\mathrm{fg}} \,,
\end{equation}
where \( r \) is the tensor-to-scalar ratio at $k = 0.05 \,{\rm Mpc}^{-1}$, \rc{} is the ratio of \causalacronym{} power to scalar power at $k = 0.01$ Mpc$^{-1}$,  ${\cal D}_\ell^{\mathrm{\causalacronym{}}}$ is the signal from Eq.~(\ref{eq:DlBB}),  and \( A_{\mathrm{LT}} \) is a free amplitude parameter that rescales the lensing \( B \)-mode template \( {\cal D}_\ell^{\mathrm{lens}} \). 
${\cal D}_\ell^{\mathrm{fg}}$ describes the foreground components in each of the likelihoods described above. 
The predicted spectra for $r = 0.02$, $\rc{} = 0.02$, and $A_{\rm LT} = 1$ are shown in Fig.~\ref{fig:models}.

We calculate posteriors using a positive-definite, uniform prior on $r$ and \rc{} over the interval \([ 0, 1]\).
The lensing amplitude is assumed to be drawn from Gaussian distribution $N(1.013, 0.023^2)$, based on the lensing measurement by \citet{qu2025unified}, to encapsulate changes to the spectrum of lensing $B$ modes from different cosmologies.

The likelihood for each dataset marginalizes over  the independent calibration and foreground parameters in each experiments' likelihood. 
An analysis with a combined foreground model is beyond the scope of this work.
All \(\Lambda\)CDM cosmological parameters are set to the \textsl{Planck} 2018 best-fit values~\cite{planck18-6}. The parameters \( r \), \( \rc{} \), and \( A_{\mathrm{LT}} \)  are sampled using Markov chain Monte Carlo with the \texttt{Cobaya} package~\citep{torrado21}.


\section{Upper limits on Early Causal Tensors}

Current CMB $B$-mode data place tight limits on the amplitude of tensor perturbations due to \causalacronym{} sources in the early universe. We jointly constrain the amplitude of power from inflationary and causal GWs while marginalizing over the lensing $B$ modes and foreground terms using the model described in Eq.~(\ref{eq:model}). The posterior distributions inferred from the combination of BICEP/Keck, SPTpol, SPT-3G, \textit{Planck}, and \textit{WMAP} data are presented in Fig.~\ref{fig:corner}. 
We find no statistically-significant detection of either \( r \) or \( \rc{}\), and the posteriors for both parameters are consistent with zero.

We obtain upper limits at 95\% confidence  of $\rc{}<$~\aclimit{} and $r <$~\rlimit{}, with $A_{\rm LT}$ tightly constrained by the prior from \citet{qu2025unified}.
Model lines at the levels of the 95\% upper limits on \rc{} and $r$ as well as the SPT-3G, SPTpol, and BK18 150 GHz bandpowers are shown in Fig.~\ref{fig:ul}. The constraint on \( \rc{}\) is driven primarily by the absence of excess power at \( 300 < \ell < 2301 \), where the signal-to-noise on the \causalacronym{} signal is expected to peak and foreground components are subdominant.

With $B$-mode data alone, the lensing and causal $B$-mode signals are significantly degenerate due to their similar angular dependence. 
The prior on lensing $B$-mode power from \cite{qu2025unified} breaks this degeneracy and strengthens the limit on \rc. As can be seen in Figure \ref{fig:corner}, $r$ and \rc{} are largely uncorrelated due to peaking at different angular scales.

\section{Upper limits on GW Backgrounds}

\begin{figure}
    \centering
    \hspace{-1cm}    \includegraphics[width=0.995\linewidth]{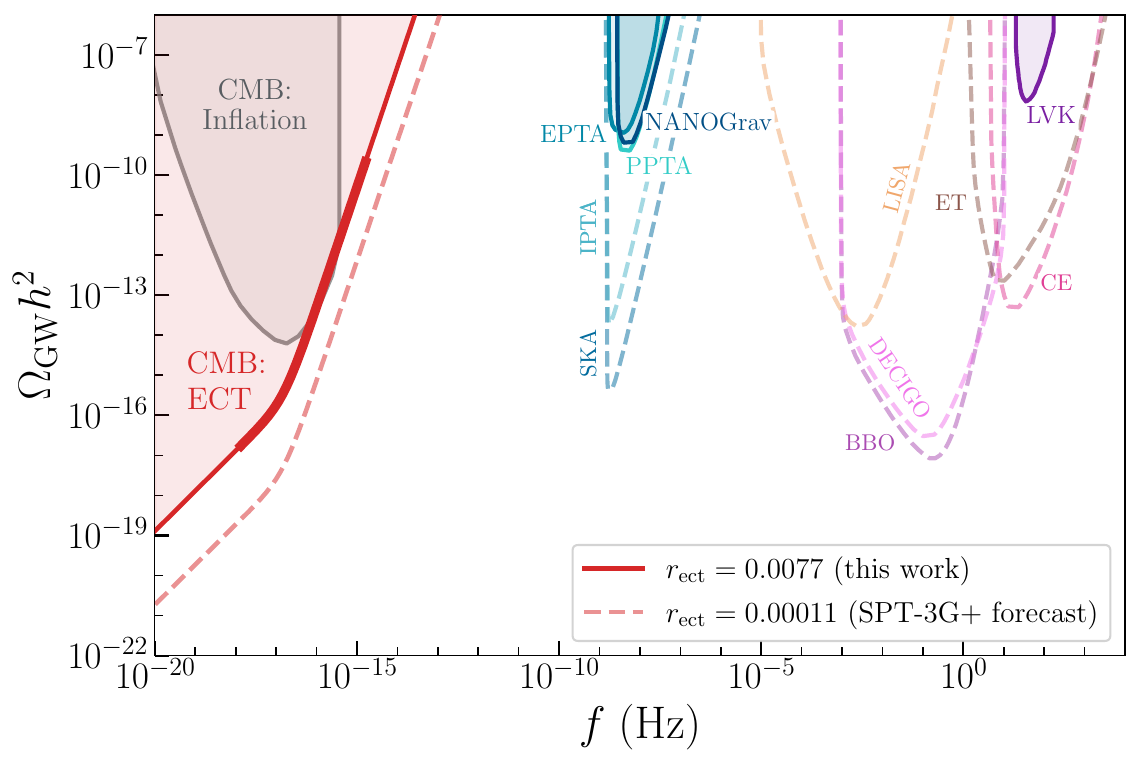}
    \caption{Limits on the energy density in GWs today, as parameterized by the spectral density $\Omega_{\rm GW}h^2$.
    The red shaded region labeled ``CMB: ECT'' is excluded at 95\% CL for GWs sourced by \causalacronym{} sources, with the bolded portion corresponding to frequencies directly probed by the CMB.
    The dashed red line is a forecast for future ECT constraints from SPT-3G+. 
    The gray shaded region labeled ``CMB: Inflation'' is excluded at 95\% CL for GWs sourced by inflation with a spectrum parameterized by $r$ and $n_t$~\cite{Lasky:2015}.
    We also show limits (solid shaded) and projected sensitivities (dashed) from various PTAs and GW interferometers \cite{Schmitz:2020syl,Caldwell:2022qsj,NANOGrav:2020,Shannon:2015,EPTA:2015qep,IPTA:2016,Weltman:2018zrl,Caprini:2019,Isoyama:2018rjb,Yagi:2011wg,ET:2019dnz,Reitze:2019iox,KAGRA:2021kbb}. Note that the red and gray exclusion curves apply \textit{only} to specific sources (\causalacronym{} sources and inflation). 
    }
    \label{fig:gw}
\end{figure}

The limit on ECT sources from CMB $B$-mode polarization measurements allows us to constrain the ECT contribution to the stochastic gravitational wave background at ultra-low frequencies well below those accessible to traditional GW probes.
We define the spectral density as
$
    \Omega_{\rm GW}  \equiv \rho^{-1}_{\rm c} d \rho_{\rm GW}/d \ln f ,
$
where $\rho_{\rm c}$ is the critical density,  $f = k/2\pi a$ is the physical frequency, and 
$
\rho_{\rm GW} = \sum_{\lambda}  \langle h_\lambda'(\tau, \vec{x})^2 \rangle/(8 \pi G a^2), 
$
is the energy density in GWs\footnote{This expression holds only at late times after the relevant modes have re-entered the horizon.}. From Eq.~\eqref{eq:Ph},  this becomes
\begin{equation}
    \Omega_{\rm GW}  = \frac{1}{\rho_{\rm c}} \frac{k^2 \mathcal{P}_h(\tau,k)}{8 \pi G a^2}  \,,
\end{equation}
where  ${\cal P}_h(\tau,k) =  \mathcal{P}_h(k) |\mathcal{T}(\tau,k)|^2$, with $\mathcal{T}$ the tensor transfer function~\cite{Turner:1993,Watanabe:2006}. The present-day spectral density is then
\begin{equation}
    \Omega_{\rm GW} h^2(f) = \frac{h^2}{3} \left( \frac{k}{H_0} \right)^2 \mathcal{P}_h(\tau_0, k) \,,
\end{equation}
where $H_0 = 100 \, h \, \text{km/s/Mpc}$, $h = 0.7$ \cite{planck18-6}, and we set $a(\tau_0) = 1$. Crucially, for modes which are superhorizon at production ($k \ll a_\star H_\star$) and re-enter during radiation domination, $\mathcal{T} \sim 1/k$, and so $\Omega_{\rm GW} h^2 \sim f^3$; for modes re-entering during matter domination, $\mathcal{T} \sim 1/k^2$ and $\Omega_{\rm GW} h^2 \sim f$. 

This scaling is manifest in Fig.~\ref{fig:gw}, where we show our limit $r_c<$\aclimit{} projected onto the $\Omega_{\rm GW} h^2$ plane (red solid). 
The red shaded region labeled ``CMB : \causalacronym{}'' is excluded at 95\% CL for GWs from \causalacronym{} sources on the basis that the corresponding $B$-mode signal would be greater than the upper limit derived here. The red dashed line indicates the projected sensitivity of SPT-3G+ \cite{anderson2022spt} (assuming 90\% delensing), corresponding to $\rc{} <$ \aclimitThreeGPlusDelensSim{}. 
We also show the inflationary CMB constraint from Ref.~\cite{Lasky:2015} (gray, shaded), which applies for GWs sourced by inflation assuming a power-law spectrum parameterized by $r$ and spectral tilt $n_t$. 
Although both the red and gray exclusions are strictly applicable only for sources with their corresponding spectral shapes $(k^0$ for inflation and $k^3$ for \causalacronym{} sources), on CMB scales, these are the only possible spectra for GWs produced before $z \sim 7 \times 10^4$. 
In Fig.~\ref{fig:gw} we also show constraints (solid curves) and projections (dashed curves) for several PTAs, current and next-generation terrestrial GW detectors, and proposed space-based interferometers. 

While model dependent, the CMB exclusion region is applicable to a wide class of early-universe scenarios. Furthermore, it cuts deep into parameter space that lies well-beyond the reach of other GW experiments, and in doing so extends our GW probes to the ultra-low frequency regime. 
This result demonstrates the power of CMB observables  to test models whose signals elude traditional GW probes.

\section{Conclusion}

In this \textit{Letter}, we have set the first-ever $B$-mode limits on causal sources of primordial GWs that predict white-noise power on CMB scales. We use a multifrequency multicomponent likelihood using BICEP/Keck, SPTpol, SPT-3G, \textit{Planck}, and \textit{WMAP} across angular scales spanning $20< \ell < 2301$. We fit the data to a model consisting of inflationary GWs, \causalacronym{} sources, lensing-induced \( B \) modes, and Galactic and extragalactic foregrounds, and find $\rc{}<$  \aclimit{} at 95\% confidence. 

Looking forward, CMB \( B \)-mode measurements can also be used to study the broader class of models that deviate from the simple \( k^3 \) white-noise template studied here. Many well-motivated scenarios
predict tensor power spectra that peak and turn over within the range of scales probed by current CMB \( B \)-mode measurements (see Fig.~\ref{fig:Ph}). 
Starting with the generic $k^3$ template allows a model-agnostic upper bound on causal GWs that can be applied to many models that generate tensor perturbations in the early universe. Future work can now build on this foundation by targeting more specific, cosmologically compelling models.


Finally, we note that our results  place world-leading limits on the lowest GW frequencies ever probed, $f \sim 10^{-19}$ Hz, assuming only that the source is non-inflationary and generates GW at redshift $z \gtrsim 7 \times 10^4$. Such frequencies are currently beyond the reach of other GW experiments, and these limits demonstrate the versatility of $B$-mode data when probing for new-physics signals.\\

\begin{acknowledgments}
\noindent {\bf Acknowledgments}: The authors would like to thank Lloyd Knox and Gil Holder for useful conversations. AI is supported by NSF Grant PHY-2310429, Simons Investigator Award No.~824870, DOE HEP QuantISED award \#100495, the Gordon and Betty Moore Foundation Grant GBMF7946, and the U.S.~Department of Energy (DOE), Office of Science, National Quantum Information Science Research Centers, Superconducting Quantum Materials and Systems Center (SQMS) under contract No.~DEAC02-07CH11359. This document was prepared using the resources of
the Fermi National Accelerator Laboratory (Fermilab),
a U.S. Department of Energy, Office of Science, Office
of High Energy Physics HEP User Facility. Fermilab
is managed by Fermi Forward Discovery Group.
CR acknowledges support from the
Australian Research Council’s Discovery Project scheme
(No. DP210102386). Support for this work for
J.Z. was provided by NASA through the NASA Hubble
Fellowship grant HF2-51500 awarded by the Space
Telescope Science Institute, which is operated by the
Association of Universities for Research in Astronomy,
Inc., for NASA, under contract NAS5-26555 and the Kavli Institute for Cosmological Physics. YT is supported by
the NSF Grant PHY-2412701. YT would also like to thank the host of Fermilab URA Visiting Scholar Program,
supported by the URA-22-F-13 fund.
\end{acknowledgments}


\bibliography{apssamp}

\end{document}